\documentclass[twocolumn,showpacs,preprintnumbers,amsmath,amssymb]{revtex4}

\usepackage{graphicx}
\usepackage{dcolumn}
\usepackage{bm}
\usepackage{subeqnar}
\usepackage{tabularx}
\usepackage{epstopdf,overpic,color}
\usepackage[ruled,noend]{algorithm2e}
\usepackage{hyperref}
\def\sech{\mathop{\mbox{\rm sech}}\nolimits}

\graphicspath{{figures/}}

\begin{document}
\preprint{APS/123-QED}

\title{Data-driven discovery of partial differential equations}
\author{Samuel H. Rudy$^1$\footnote{Electronic address: \texttt{shrudy@uw.edu}}, Steven L. Brunton$^2$, Joshua L. Proctor$^3$, 
and J. Nathan Kutz$^1$}
%
\affiliation{$^1$ Department of Applied Mathematics, University of Washington, Seattle, WA. 98195}
\affiliation{$^2$ Department of Mechanical Engineering, University of Washington, Seattle, WA. 98195}
\affiliation{$^3$ Institute for Disease Modeling, , 3150 139th Ave SE, Bellevue, WA 98005}
\date{\today}
\begin{abstract}
We propose a sparse regression method capable of discovering the governing partial differential equation(s)
of a given system by time series measurements in the spatial domain.  
The regression framework relies on sparsity promoting techniques to select the nonlinear and partial derivative terms terms of the governing equations that most accurately represent the data, bypassing a combinatorially large search through all possible candidate models.   
The method balances model complexity and regression accuracy by selecting a parsimonious model via Pareto analysis.  Time series measurements can be made in an 
Eulerian framework where the sensors are fixed spatially, or in a Lagrangian framework where the sensors move with the dynamics.
The method is computationally efficient, robust, and demonstrated to work on a variety of canonical problems of mathematical
physics including Navier-Stokes, the quantum harmonic oscillator, and the diffusion equation.  Moreover, the method is capable of disambiguating between potentially non-unique dynamical terms by using multiple time series taken with different initial data.  
Thus for a traveling wave, the method can distinguish between a linear wave equation or the Korteweg-deVries equation, for instance.   
The method provides a promising new technique for discovering governing equations
and physical laws in parametrized spatio-temporal systems where first-principles derivations are intractable. 
\end{abstract}

\maketitle

Data-driven discovery methods, which have been enabled in the last decade by the plummeting cost of sensors, data storage and computational resources,  are having a transformative impact on the sciences, enabling a variety of innovations for characterizing high dimensional data generated from experiments.  
Less well understood is how to uncover underlying physical laws and/or governing equations from time series data that exhibit spatio-temporal activity.  
Traditional theoretical methods for deriving the underlying partial differential equations (PDEs) are rooted in conservation laws, physical principles and/or phenomenological behaviors.  These first principle derivations lead to many of the canonical models of mathematical physics.  However, there remain many complex systems that have eluded quantitative analytic descriptions or even characterization of a suitable choice of variables (e.g. neuroscience, power grid, epidemiology, finance, ecology, etc).  We propose an alternative method to derive governing equations based solely on time series data collected at a fixed number of spatial locations.  
Using innovations in sparse regression, we discover the terms of the governing PDE that most accurately represent the 
data from a large library of potential candidate functions.  
Importantly, measurements can be made in an Eulerian framework where the sensors are fixed spatially, or in a Lagrangian framework
where the sensors move with the dynamics.
We demonstrate the success of the method by deriving, from time series data alone, many canonical models of
mathematical physics.

Methods for data-driven discovery of dynamical systems~\cite{Crutchfield1987cs} include equation-free modeling~\cite{Kevrekidis2003cms}, empirical dynamic modeling~\cite{Sugihara2012science,Ye2015pnas}, modeling emergent behavior~\cite{Roberts2014book}, and automated inference of dynamics~\cite{Schmidt2011pb,Daniels2015naturecomm,Daniels2015plosone}.  
In this series of developments, seminal contributions leveraging symbolic regression and an evolutionary algorithm~\cite{Bongard2007pnas,Schmidt2009science} were capable of directly determining nonlinear dynamical system from data.   
More recently, sparsity promoting techniques~\cite{Tibshirani1996lasso} have
been used to robustly determine, in a highly efficient computational manner, 
the governing dynamical system~\cite{Brunton2016,Mangan2016}. 
%
%
Both the evolutionary~\cite{Schmidt2009science} and sparse~\cite{Brunton2016} symbolic regression methods avoid overfitting by selecting parsimonious models that balance model accuracy
with complexity via Pareto analysis.  

\begin{figure*}[t]
 \center
  \includegraphics[width=0.9\textwidth]{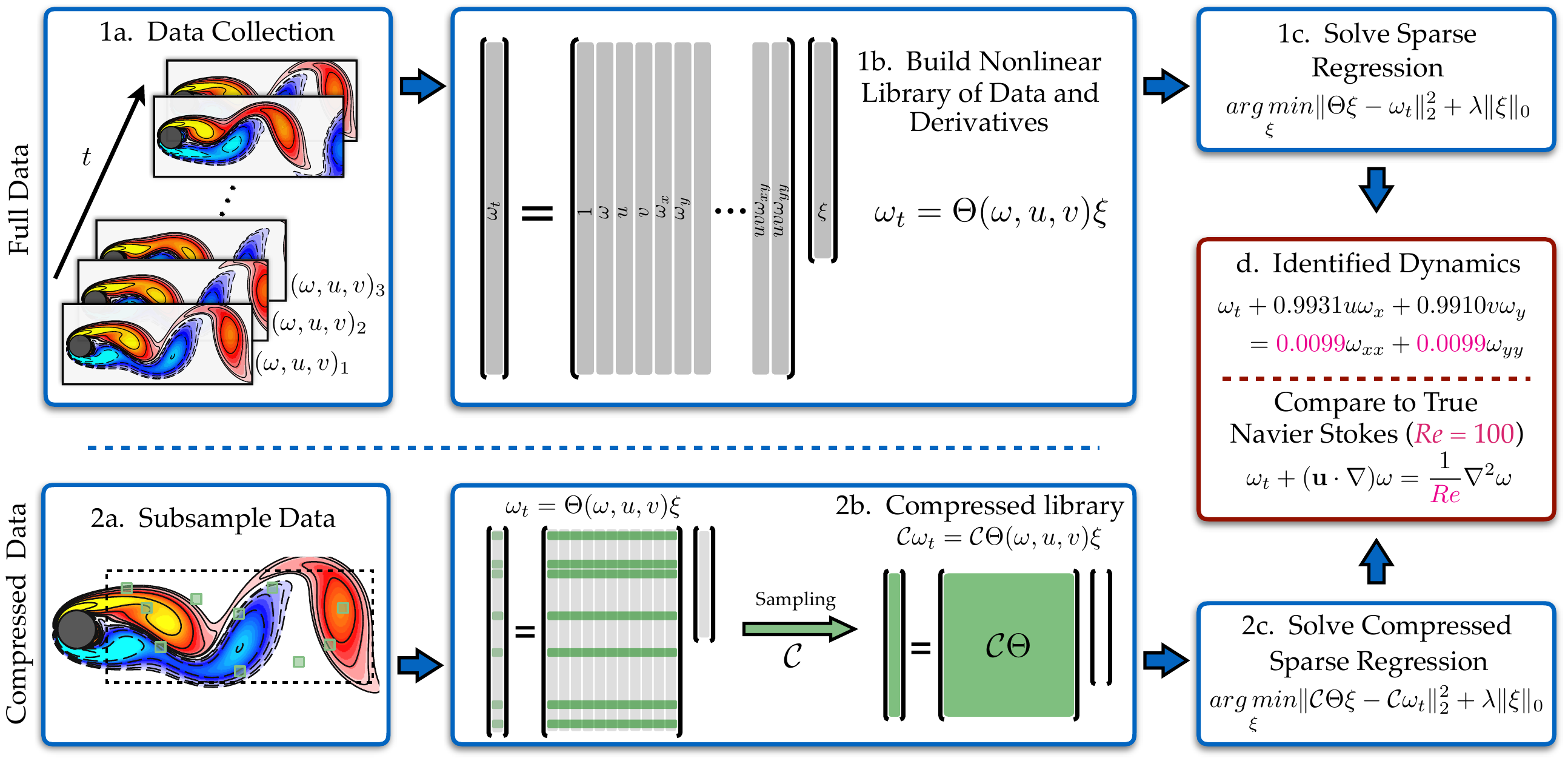}
  \vspace{-.15in}
  \caption{Steps in the PDE functional identification of nonlinear dynamics (PDE-FIND) algorithm, applied to infer the Navier-Stokes equation from data.  \textbf{1a.}  Data is collected as snapshots of a solution to a PDE. \textbf{1b.}  Numerical derivatives are taken and data is compiled into a large matrix $\mathbf{\Theta}$, incorporating candidate terms for the PDE. \textbf{1c.}  Sparse regressions is used to identify active terms in the PDE. \textbf{2a.}  For large datasets, sparse sampling may be used to reduce the size of the problem.  \textbf{2b.} Subsampling the dataset is equivalent to taking a subset of rows from the linear system in \eqref{Eq:PDEFIND}.  \textbf{2c.} An identical sparse regression problem is formed but with fewer rows.  \textbf{d.}  Active terms in $\xi$ are synthesized into a PDE.}
  \vspace{-.15in}
  \label{nav_stoks}
\end{figure*}

The method we present is able to select linear and/or nonlinear terms, including spatial derivatives, resulting in the identification of PDEs from data.  
Previous methods are able to identify ODEs from data, but not partial derivative terms~\cite{Brunton2016}.  
Only those terms that are most informative about the dynamics are selected as part of the discovered PDE.  The generalization presented here is critically important since the majority of canonical models in mathematical physics contain spatio-temporal dynamics.  The addition of spatial structure is nontrivial and requires modification of the methodology and data collection (Eulerian or Lagrangian measurements) in order to circumvent potential ambiguities.  The resulting algorithm, {\em PDE functional identification of nonlinear dynamics} (PDE-FIND), is applied to numerous canonical models of mathematical physics.  

In what follows, we consider a PDE of the form
\begin{equation}
  u_t = N (u, u_x, u_{xx}, \cdots, x, \mu)
\end{equation}
where the subscripts denote partial differentiation in either time or space, and $N(\cdot)$ is
an unknown right-hand side that is generally a nonlinear function of $u(x,t)$, its
derivatives, and parameters in $\mu$.  Our objective is to construct $N(\cdot)$ given time series measurements of the
system at a fixed number of spatial locations in $x$.  A key assumption is that the function 
$N(\cdot)$ consists of only a few terms, making it sparse in the space of possible functions.
As an example, Burgers' equation ($N=- uu_x + \mu u_{xx}$) and the harmonic oscillator
($N=-i \mu x^2 u - i \hbar u_{xx}/2$)  each have two terms. 
Thus sparse regression allows
one to determine \emph{which} right hand side terms are non-zero without an intractable ($np$-hard) combinatorial brute-force search.   

The sparse regression and discovery method (See Fig.~\ref{nav_stoks}) begins by first collecting all the spatial, time series data into a single column vector ${\bf U}\in\mathbb{C}^{m n}$ representing data collected over $m$ time points and $n$ spatial locations.  We also consider any additional input such as a known potential for the Schr\"odinger equation, or the magnitude of complex data, in a column vector ${\bf Q}\in\mathbb{C}^{m n}$. 
Next, a library ${\bf\Theta} ({\bf U}, {\bf Q}) \in \mathbb{C}^{mn \times D}$ of $D$ candidate linear and nonlinear terms and partial derivatives for the PDE is constructed.  Each column of ${\bf\Theta} ({\bf U}, {\bf Q})$ lies in $\mathbb{C}^{mn}$ and contains the values of a candidate term in the PDE across all gridpoints on which data is collected, as illustrated in Fig.~\ref{nav_stoks}.  For example, a column of ${\bf \Theta} ({\bf U}, {\bf Q})$ may be $q^2u_{xx}$.
The  PDE in this library is:
\begin{equation}
{\bf U}_t = {\bf \Theta} ({\bf U}, {\bf Q}) \xi.\label{Eq:PDEFIND}
\end{equation}
Each entry in $\xi$ is a coefficient corresponding to a term in the PDE, and for canonical PDEs, the vector $\xi$ is \emph{sparse}, meaning that only a few terms are active. 

Proper evaluation of the numerical derivatives is the most challenging and critical task for the success of the method.  Given the well-known accuracy problems with finite-difference approximations, we instead use polynomial interpolation for differentiating noisy data.  The method depends on the degree of polynomial and number of points used.  In some cases, filtering the noise via the singular value decomposition is necessary (See Supplementary Materials for details).

In general, we require the sparsest vector $\xi$ that satisfies \eqref{Eq:PDEFIND} with a small residual.  Instead of an intractable combinatorial search through all possible sparse vector structures, a common technique is to relax the problem to a convex $\ell_1$ regularized least squares~\cite{Tibshirani1996lasso}; however, this tends to perform poorly with highly correlated data.  Instead, we approximate the problem using candidate solutions to a ridge regression problem with hard thresholding, which we call sequential threshold ridge regression (STRidge in Algorithm 1).  For a given tolerance and $\lambda$, this gives a sparse approximation to $\xi$.  
\begin{algorithm}[t]
    \caption{STRidge($\mathbf{\Theta}, \mathbf{U}_t, \lambda, tol, \text{iters}$)}
    \label{alg1}   
    $\hat{\xi} = arg\,min_{\xi} \|\mathbf{\Theta} \xi - \mathbf{U}_t\|_2^2 + \lambda \|\xi \|_2^2$\hspace{8 mm}\# ridge regression\\
	bigcoeffs = $\{ j : |\hat{\xi}_j| \geq tol \}$  \hspace{6 mm} \# select large coefficients\\
    $\hat{\xi}$[ $\sim$ bigcoeffs] = 0 \hspace{19 mm} \# apply hard threshold\\
    $\hat{\xi}$[bigcoeffs] = STRidge($\mathbf{\Theta}[:, \text{bigcoeffs}], \mathbf{U}_t, tol, \text{iters}-1$)\\
    \hspace{26 mm} \# recursive call with fewer coefficients\\
    return $\hat{\xi}$
\end{algorithm}
We iteratively refine the tolerance of Algorithm 1 to find the best predictor based on the selection criteria, \begin{equation}
\hat{\xi} = \text{argmin}_{\xi}\|\mathbf{\Theta}(\mathbf{U},\mathbf{Q})\xi-\mathbf{U}_t\|_2^2 + \epsilon \kappa(\mathbf{\Theta}(\mathbf{U},\mathbf{Q}))\|\mathbf{\xi}\|_0
\end{equation}
where $\kappa(\mathbf{\Theta})$ is the condition number of the matrix $\mathbf{\Theta}$, indicating stronger regularization for ill-posed problems.  Penalizing $\|\xi\|_0$ discourages over fitting by selecting from the optimal position in a Pareto front.

\begin{figure}[t!]
 \center
  \includegraphics[width=0.38\textwidth]{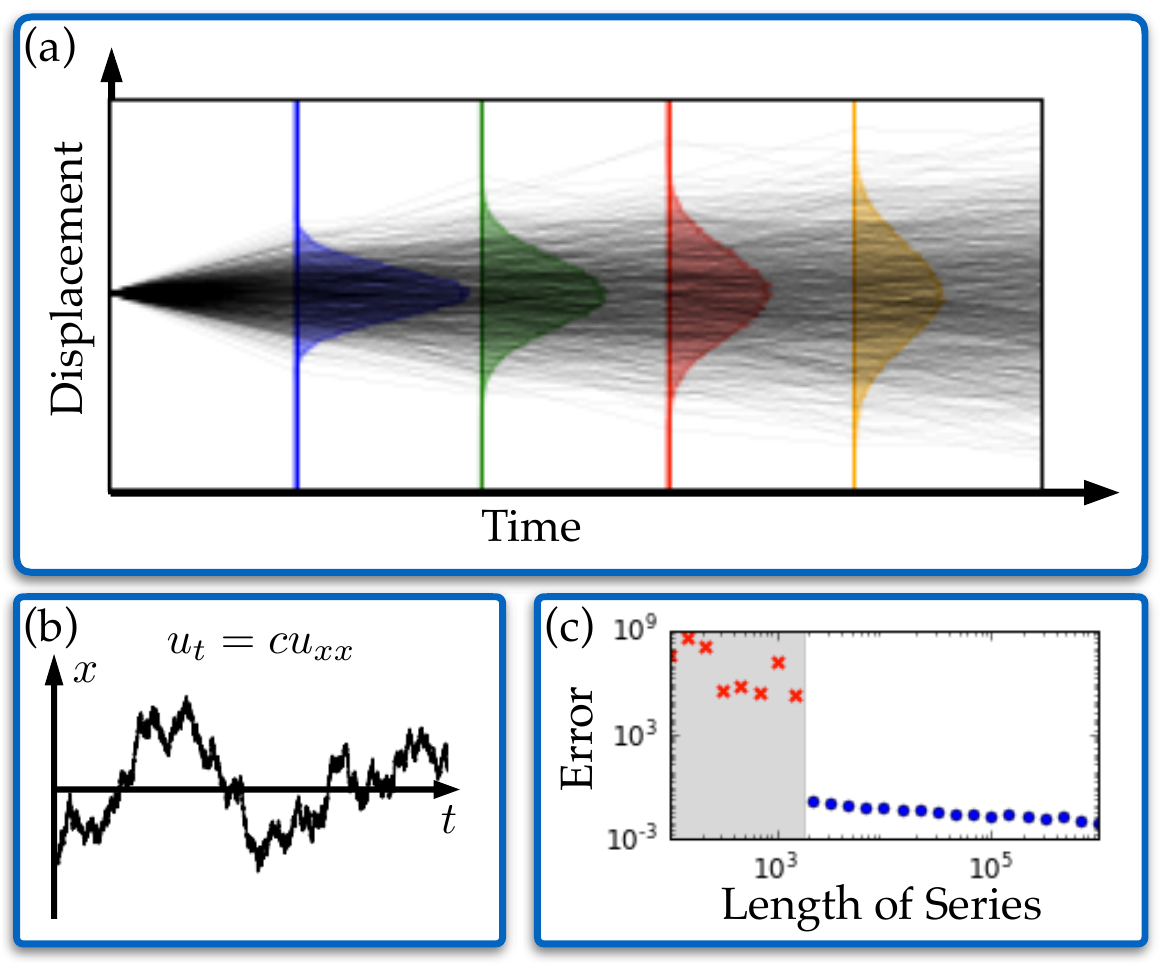} 
  \vspace{-.1in}
    \caption{Inferring the diffusion equation from a single Brownian motion.  (\textbf{a})  Time series is broken into many short random walks that are used to construct histograms of the displacement.  (\textbf{b})  The Brownian motion trajectory, following the diffusion equation.  (\textbf{c})  Parameter error ($\|\xi^* - \hat{\xi}\|_1$) vs. length of known time series.  Blue symbols correspond to correct identification of the structure of the diffusion model, $u_t=cu_{xx}$.}
  \label{heat}
  \vspace{-.1in}
  \end{figure}

PDE-FIND differs from previous sparse identification algorithms~\cite{Brunton2016}, where high-dimensional data from a PDE is handled by first applying dimensionality reduction, such as proper orthogonal decomposition (POD), to obtain a few dominant coherent structures in the data.  Traditionally, an ODE is then identified on the coefficients of these energetic modes, resulting in a model that resembles a Galerkin projection onto POD modes~\cite{HLBR_turb}. In contrast, the PDE-FIND algorithm directly identifies the fewest terms required to balance the governing PDE. 

As a first demonstration of the method, we consider 
one of the fundamental results of the early 20th century concerning the relationship between random 
walks (Brownian motion) and diffusion.  The theoretical connection between these two was first made
by Einstein in 1905~\cite{Einstein1905brownian} and was part of the {\em Annus Mirablis papers} which lay the foundations of modern
physics.   We use the method proposed here to sample the movement of a random walker, which is
effectively a Lagrangian measurement coordinate, in order to verify that it can reproduce the well-known diffusion equation.  By biasing the random walk, we can also produce the generalization
of advection-diffusion in one-dimension.  Figure~\ref{heat} shows the success of the method in identifying the correct diffusion model from a random walk trajectory.  Given a sufficiently long time series with high enough resolution, the method produces the heat equation for the evolution of the probability distribution function.
Thus a PDE is derived from a single time series representing discrete measurements of a continuous stochastic process.  Specifically, a single time series is broken into pieces to construct a histogram approximating the distribution function of a trajectory's future position at various timesteps.  The resulting function is fit to a PDE using PDE-FIND, thus allowing us to sample Brownian motion in order to derive the diffusion equation.

\begin{figure}
\begin{center}
  \includegraphics[width=0.38\textwidth]{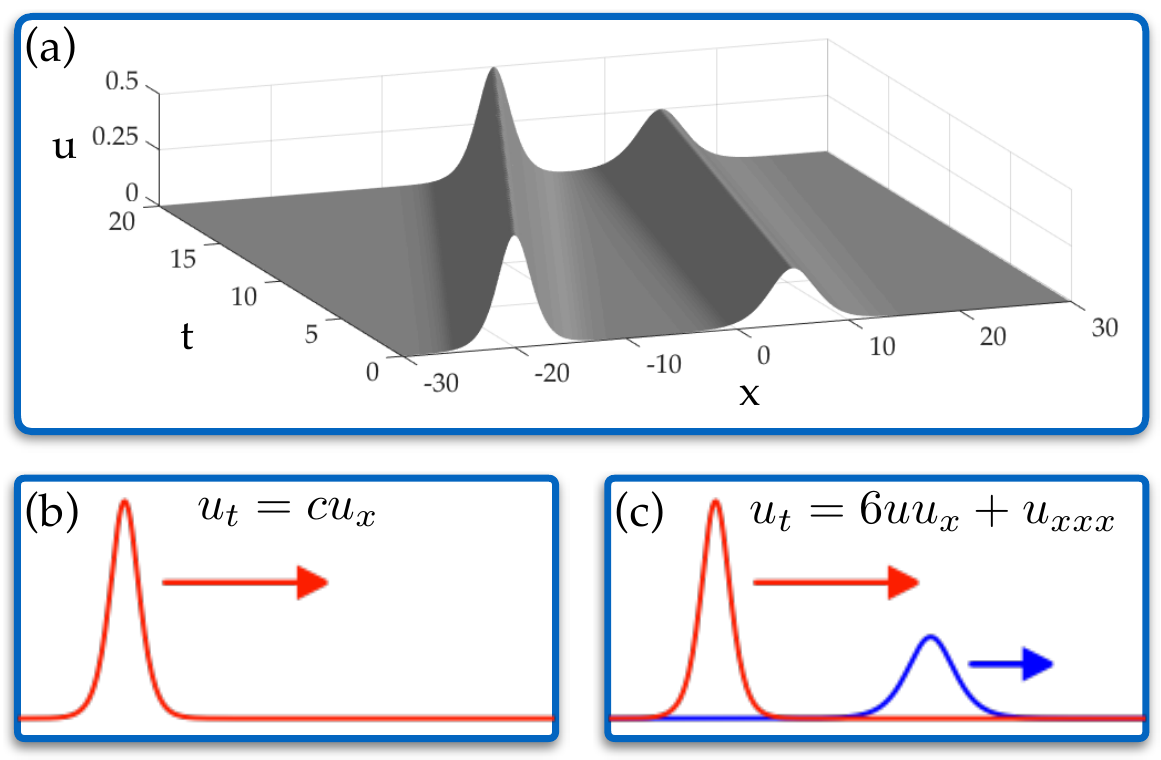} 
  \vspace{-.1in}
  \caption{Inferring nonlinearity via observing solutions at multiple amplitudes.  (\textbf{a})  An example 2-soliton solution to the KdV equation.  (\textbf{b})  Applying our method to a single soliton solution determines that it solves the standard advection equation.  (\textbf{c}) Looking at two completely separate solutions reveals nonlinearity.    \label{kdv_advection}}
  \end{center}
  \vspace{-.2in}
\end{figure}

{\small 
\begin{table*}[t]
\begin{center}
    \begin{tabular}{ | l | l | l | p{5.5cm} |}
    \hline
    PDE & Form & Error (no noise, noise) & Discretization\\ \hline
      \raisebox{-0.12in}{\includegraphics[width=0.1\textwidth, height=11mm]{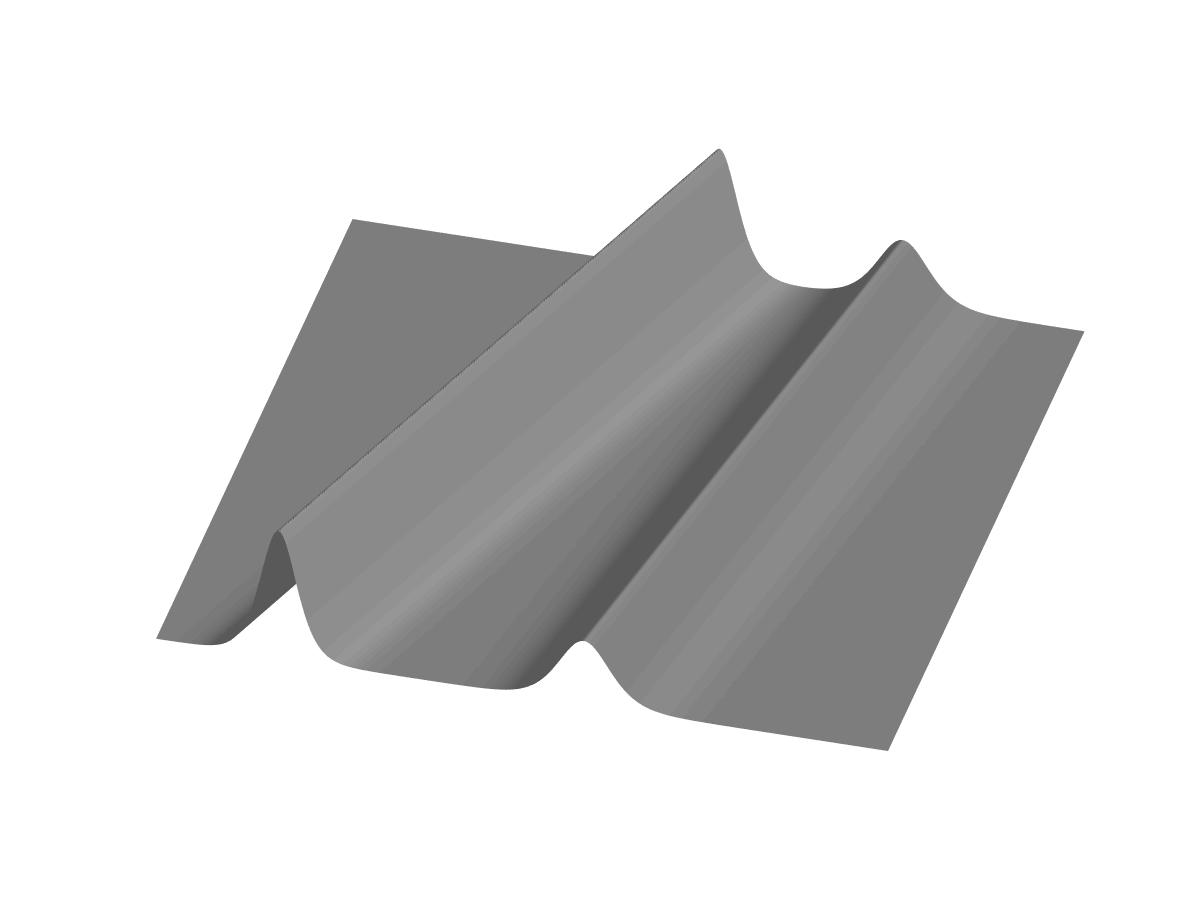} }
   \!\!\!\!  KdV & $u_t + 6 uu_x + u_{xxx} = 0$ 
     & 
       $1\% \!\pm\! 0.2\%,
         7\% \!\pm\! 5\% $
     &  
          $x \!\!\in\!\![-30,30], n\!\!=\!\!512$,  
          $t \!\!\in\!\![0,20], m\!\!=\!\!201$    \\ \hline 
  \raisebox{-0.12in}{\includegraphics[width=0.1\textwidth, height=11mm]{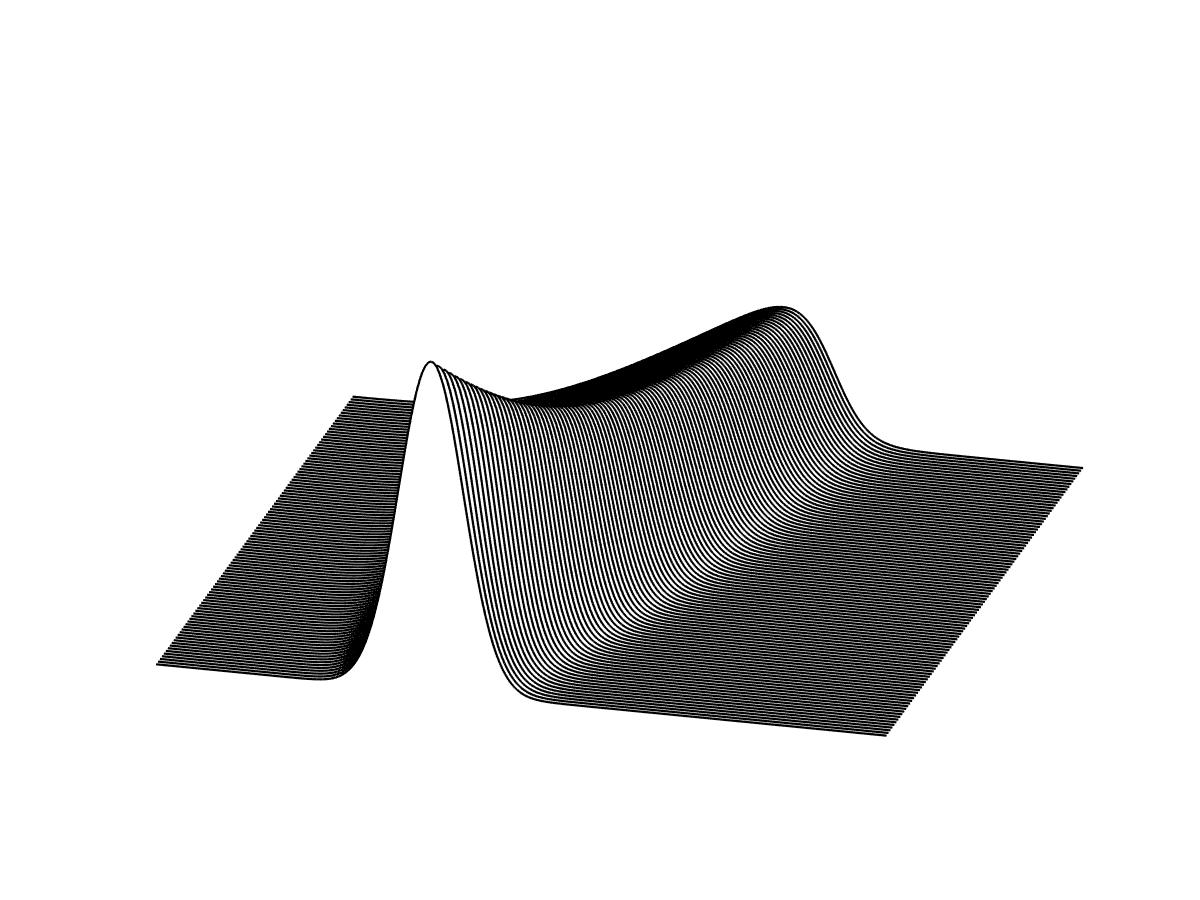} }
  \!\!\!\! Burgers & $u_t + uu_x -\epsilon u_{xx}= 0$ 
   &
    $0.15\% \!\pm\! 0.06\%,
         0.8\% \!\pm\! 0.6\% $
   &
   $x \!\!\in\!\![-8,8], n\!\!=\!\!256$,  
          $t \!\!\in\!\![0,10], m\!\!=\!\!101$
           \\ \hline
   \raisebox{-0.12in}{\includegraphics[width=0.1\textwidth, height=11mm]{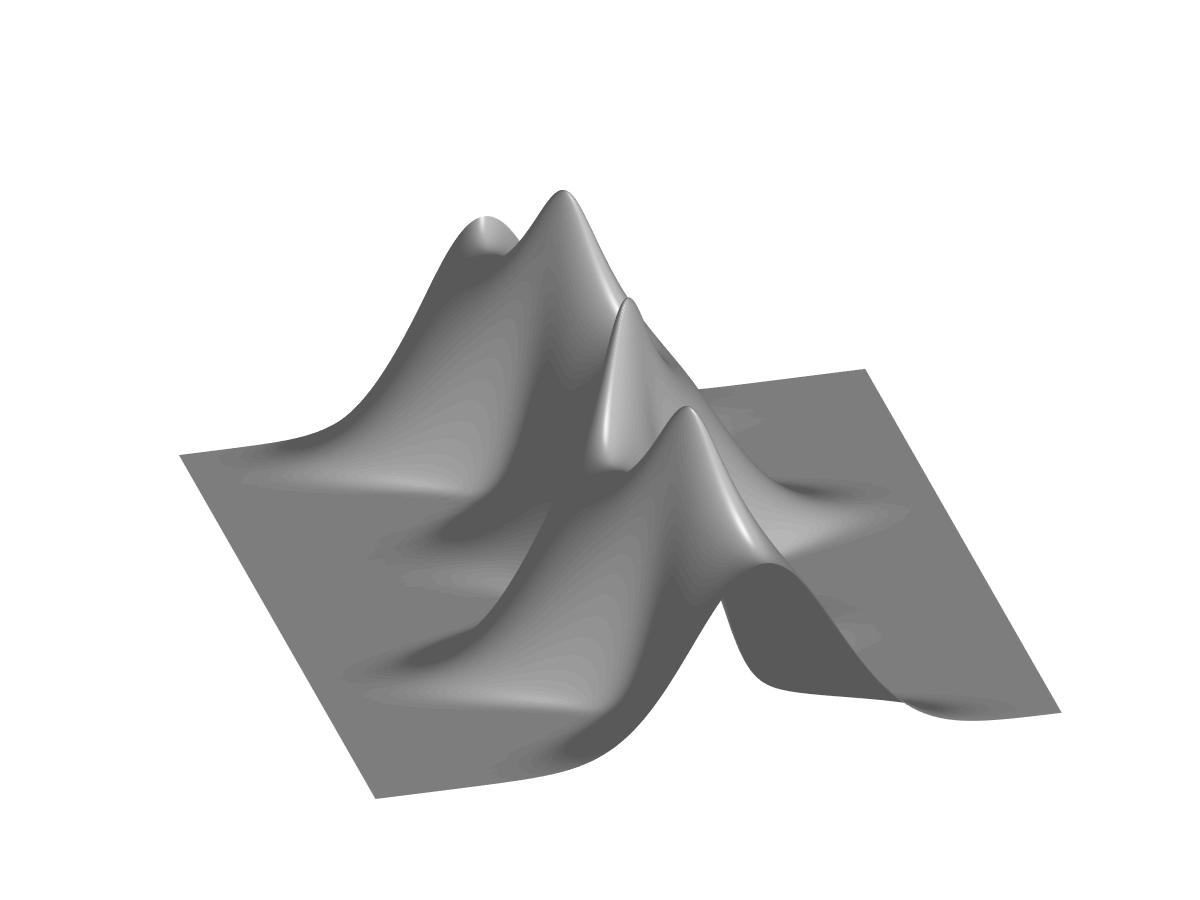} }  \!\!\!\!Schr\"odinger & $i u_t + \frac{1}{2} u_{xx} - \frac{x^2}{2} u =0$ 
   &
   $0.25\% \!\pm\! 0.01\%,
         10\% \!\pm\! 7\% $
    &
   $x \!\!\in\!\![-7.5,7.5], n\!\!=\!\!512$,  
          $t \!\!\in\!\![0,10], m\!\!=\!\!401$
           \\ \hline
  \raisebox{-0.12in}{\includegraphics[width=0.1\textwidth, height=11mm]{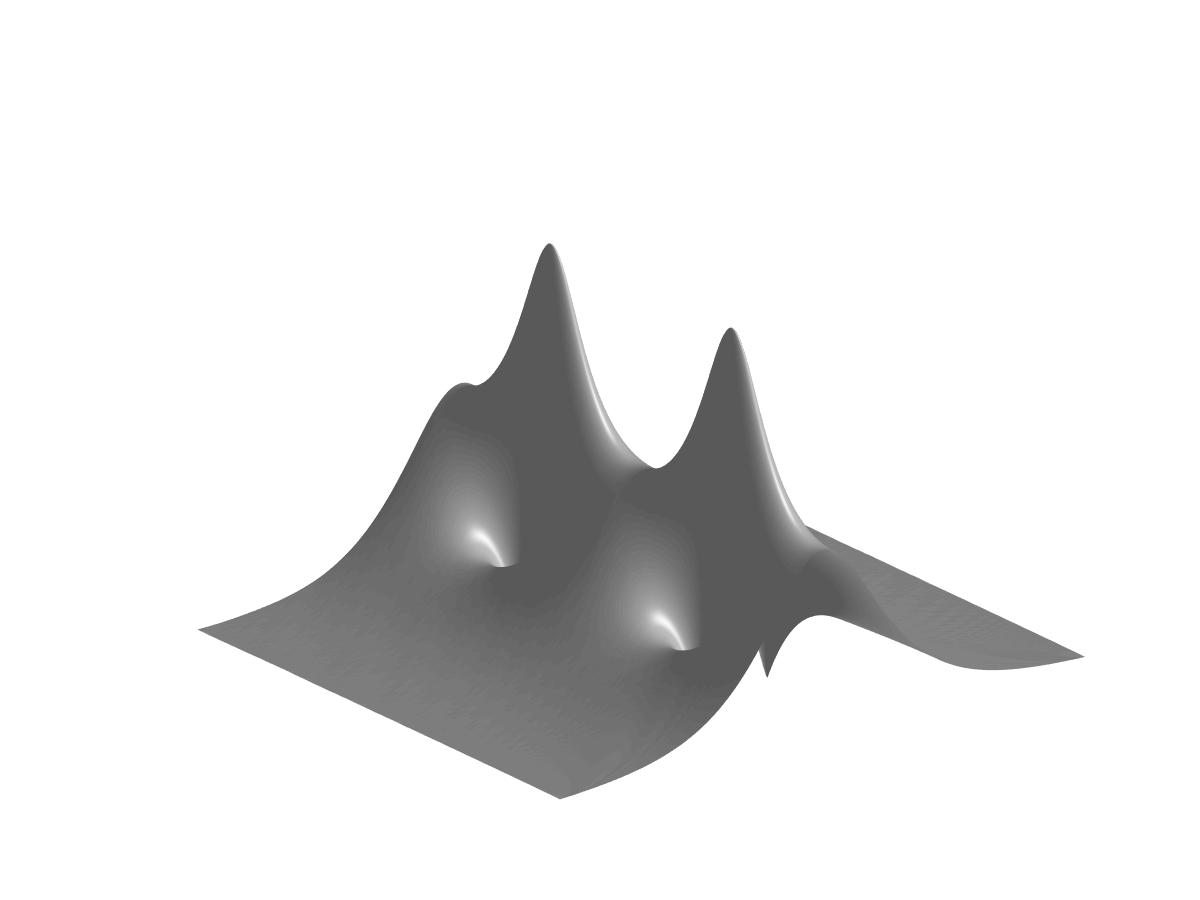} }
     \!\!\!\!NLS & $iu_t + \frac{1}{2}u_{xx} + |u|^2u = 0$ 
     &
     $0.05\% \!\pm\! 0.01\%,
         3\% \!\pm\! 1\% $
     &
     $x \!\!\in\!\![-5,5], n\!\!=\!\!512$,  
          $t \!\!\in\!\![0,\pi], m\!\!=\!\!501$
           \\ \hline
  \raisebox{-0.12in}{\includegraphics[width=0.1\textwidth, height=11mm]{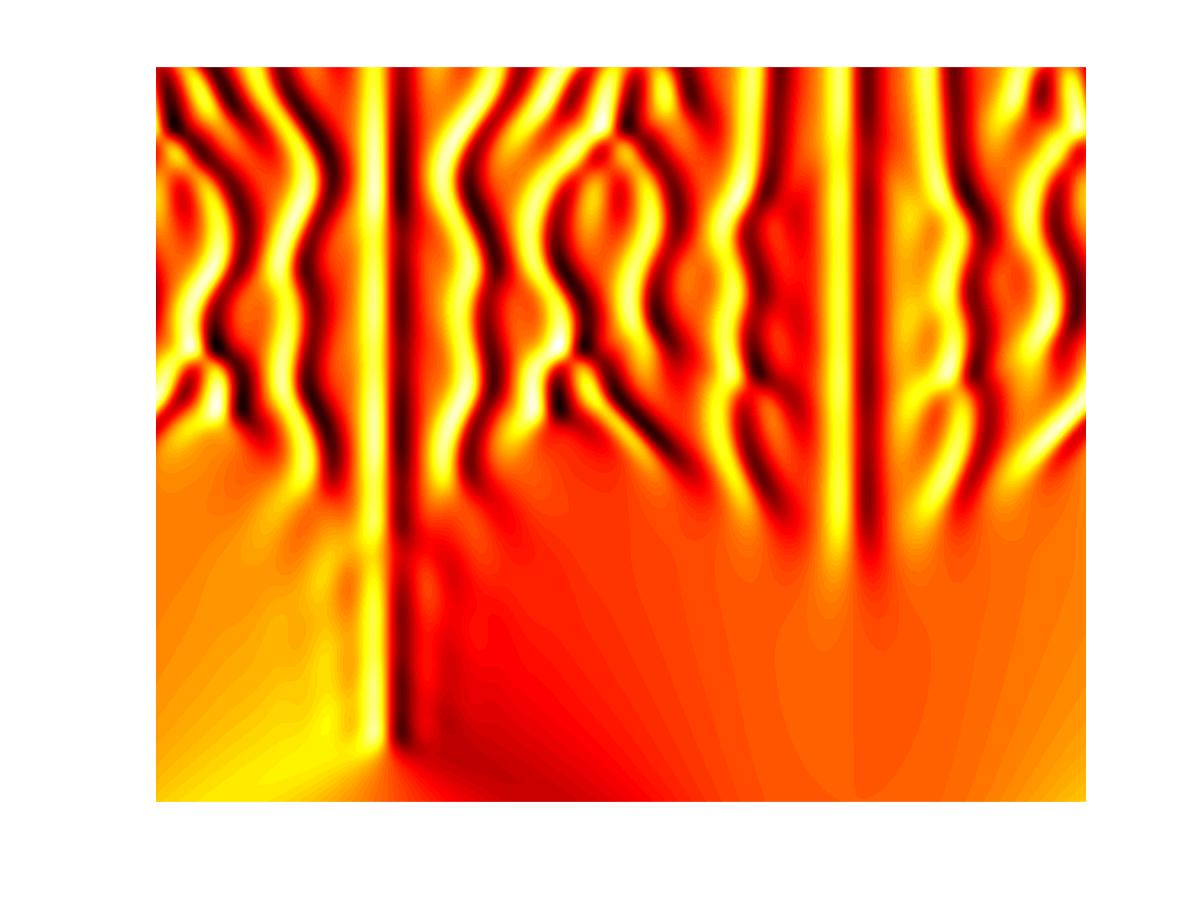} }
    \!\!\!\!KS & $ u_t + uu_x + u_{xx} + u_{xxxx} =0 $  
    &
    $1.3\% \!\pm\! 1.3\%,
         70\% \!\pm\! 27\% $
     &
    $x \!\!\in\!\![0,100], n\!\!=\!\!1024$,  
          $t \!\!\in\!\![0,100], m\!\!=\!\!251$
           \\ \hline
    \raisebox{-0.12in}{\begin{overpic}[width=0.1\textwidth, height=8mm]{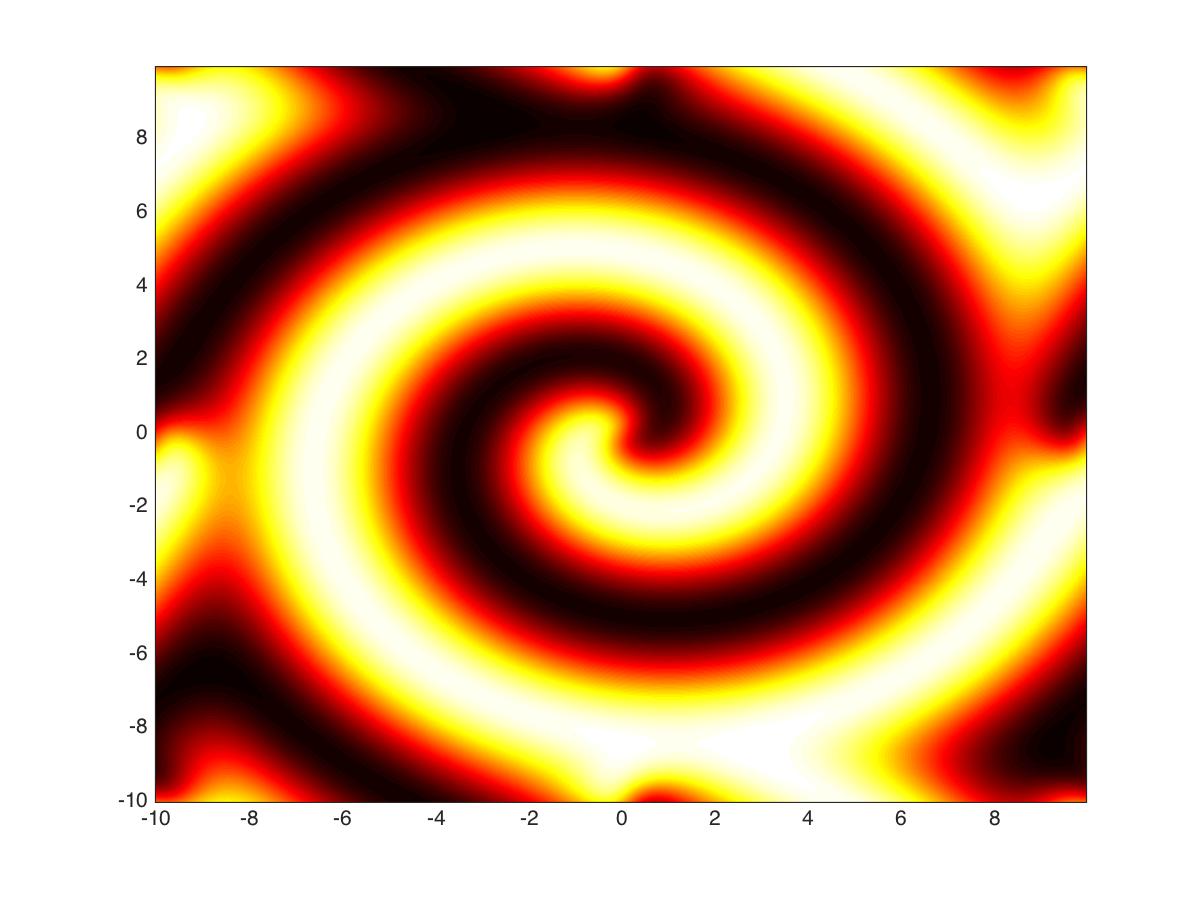} 
    \put(0,20){$u$}
    \end{overpic} } \!\!\!\!Reaction  &  $u_t = 0.1 \nabla^2 u+ \lambda (A) u - \omega(A) v$ & & \\[-0.1in]  \hspace*{.65in} Diffusion  &  
            $v_t = 0.1 \nabla^2 v + \omega (A) u + \lambda (A) v $ 
             &
             $0.02 \% \pm 0.01 \%$, $3.8 \% \pm 2.4 \%$
             &
             $x,y \!\!\in\!\![-10,10], n\!\!=\!\!256$,  
          $t \!\!\in\!\![0,10], m\!\!=\!\!201$ 
          	 \\[-.1in]
        \raisebox{-0.12in}{\begin{overpic}[width=0.1\textwidth, height=8mm]{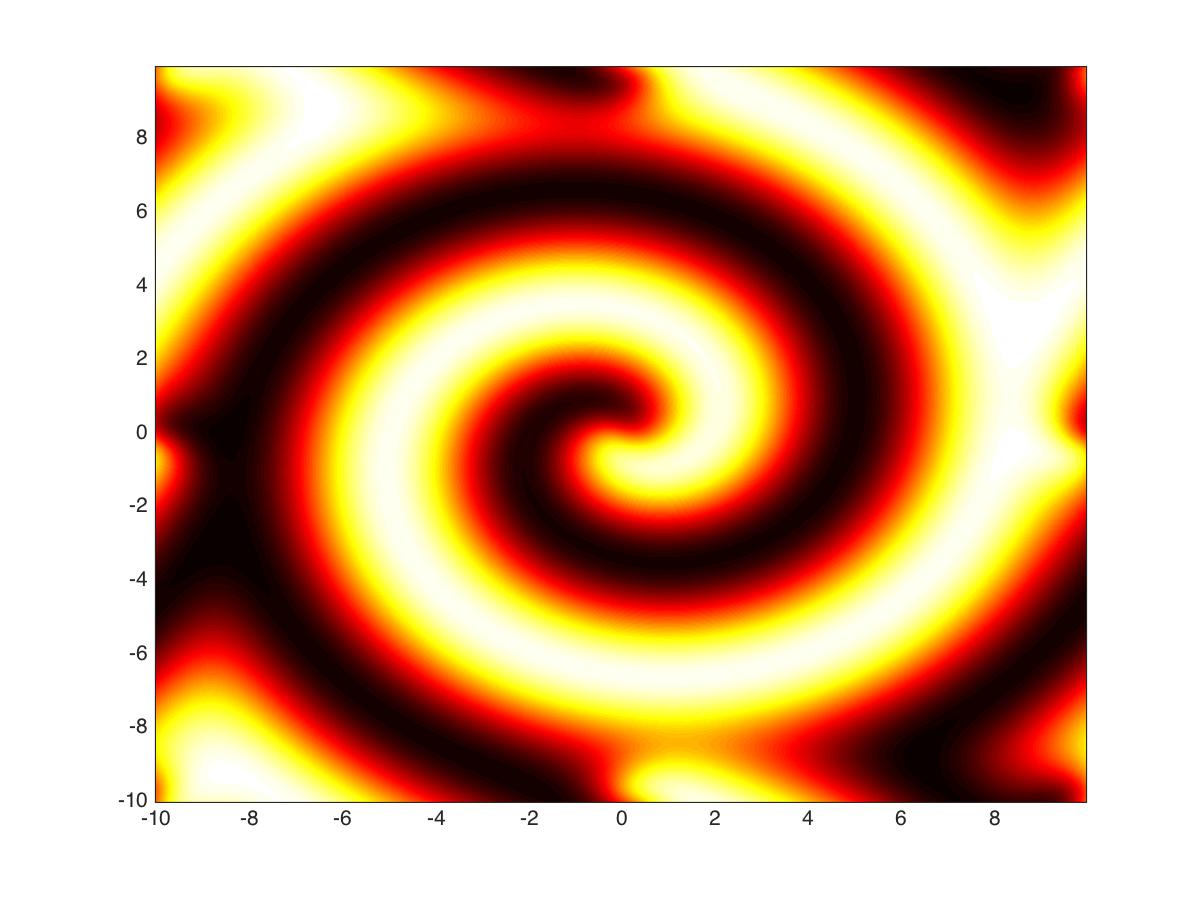}\put(0,20){$v$}
    \end{overpic} }         & $A^2\!=\!u^2\!+\!v^2 \!,  \omega \!\!=\!-\beta A^2  \!, \lambda  \!\!=\!\! 1\!-\! A^2 $ &   & subsample $1.14\%$ \! 
           \\ \hline
\raisebox{-0.12in}{\includegraphics[width=0.08\textwidth, height=9mm]{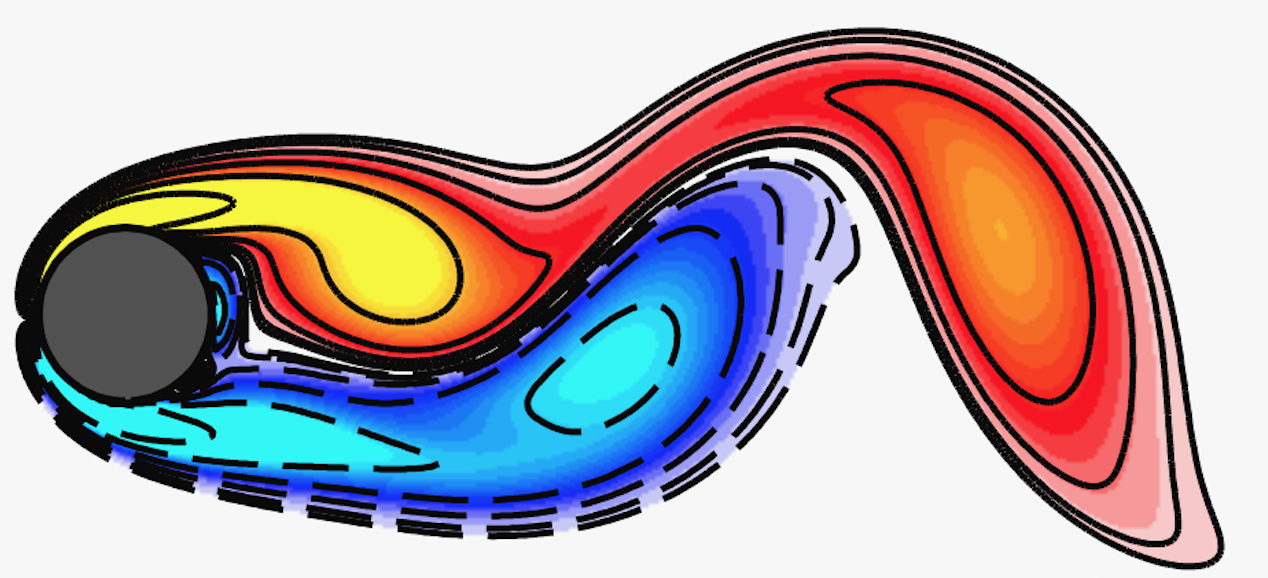} } \!\!\!\! Navier Stokes &
  $\omega_t + ( {\bf u}\cdot \nabla ) \omega = \frac{1}{Re} \nabla^2 \omega$  
  &
   $1 \% \pm 0.2 \%$ , $7 \% \pm 6 \%$
   & 
  			$x\!\!\in\!\![0,9], n_x\!\!=\!\!449$, 
        	$y\!\!\in\!\![0,4], n_y\!\!=\!\!199$, 
          	\newline
          	$t \!\!\in\!\![0,30], m\!\!=\!\!151$, subsample $2.22\%$ \! 
          \\ 
	 \hline
    \end{tabular}
\end{center}
\vspace*{-.2in}
\caption{Summary of regression results for a wide range of canonical modes of mathematical physics.  In each example, the correct model structure is identified using PDE-FIND.  The spatial and temporal sampling used for the regression is given along with the error produced in the parameters of the model for both no noise and 1\% noise.  In the reaction-diffusion (RD) system, 0.5\% noise is used. For Navier Stokes and Reaction Diffusion, the percent of data used in subsampling is also given.} 
\label{ta:pdes}
\end{table*}
}

A second canonical example is the KdV equation modeling the unidirectional propagation of small-amplitude, long water waves or shallow-water waves.  Discovered first by Boussinesq in 1877 and later developed by Korteweg and deVries in 1895,
it was one of the earliest models known to have soliton solutions.   One potential viewpoint of the equation
is as a dispersive regularization of Burgers' equation.  The KdV evolution is given by
\begin{equation}
  u_t + 6 u u_x + u_{xxx} = 0 \, ,
  \label{eq:kdv}
\end{equation}
with soliton solutions taking the form $u(x,t) \!=\! ({c}/{2}) \sech^2 \! \left[ ({\sqrt{c}}/{2}) (x\!-\! ct \!-\! x_0) \right]$. 
These solutions propagate with a speed proportional to their amplitude $c$.  Interestingly, if one observes a single
propagating soliton, it would be indistinguishable from a solution to the one-way wave equation $u_t +c u_x =0$.  As such, it presents a challenge to the sparse regression framework as the sparsity promotion would select
the one-way wave equation over the KdV equation since it has the sparsest representation.  This ambiguity in the
governing PDE is rectified by constructing time series data for more than a single initial amplitude.  
Figure~\ref{kdv_advection} demonstrates the evolution of two KdV solitons of differing amplitudes, which allows for uniquely determining the governing PDE (\ref{eq:kdv}).

Table~\ref{ta:pdes} applies the methodology proposed to a wide range of canonical models from mathematical physics.  The PDEs selected represent a wide range of physical systems, displaying both Hamiltonian (conservative) dynamics and dissipative nonlinear dynamics along with periodic to chaotic behavior.  Aside from the quantum oscillator (3rd row), all the dynamics observed are strongly nonlinear.  Remarkably, the method is able to discover each physical system even if significantly subsampled spatially.  The space and time sampling required, along with the accuracy in recovering the PDE parameters with and without noise, are detailed in the Table.   This highlights the broad applicability of the method and the success of the technique in discovering governing PDEs. 

PDE-FIND is a viable, data-driven tool for modern applications where first-principles derivations may be intractable (e.g. neuroscience, epidemiology, dynamical networks), but where new data recordings and sensor technologies are revolutionizing our understanding of physical and/or biophysical processes on spatial domains.  To our knowledge, this is the first data-driven regression technique that explicitly accounts for spatial derivatives in discovering physical laws, thus allowing for a regression to an operator on an infinite-dimensional space.  The ability to discover physical laws instead of approximate, low-dimensional subspaces enables significantly improved future state predictions as well as the discovery of parametric dependencies.  For instance, we can discover the Navier Stokes equation at $Re=100$ and use this knowledge to accurately simulate a fully turbulent system at $Re=10000$ where no data was collected.  This represents a significant paradigm shift when compared with most data-driven, machine learning architectures where accurate predictions can only be made near parameter regimes where the data was sampled. 

\noindent\footnotesize{Code and supplementary material:\\
\href{https://github.com/snagcliffs/PDE-FIND}{https://github.com/snagcliffs/PDE-FIND}
}

\setlength{\bibsep}{1.75pt}
\footnotesize{\bibliographystyle{unsrt}
\bibliography{references}} 

\begin{thebibliography}{15}
\expandafter\ifx\csname natexlab\endcsname\relax\def\natexlab#1{#1}\fi
\expandafter\ifx\csname bibnamefont\endcsname\relax
  \def\bibnamefont#1{#1}\fi
\expandafter\ifx\csname bibfnamefont\endcsname\relax
  \def\bibfnamefont#1{#1}\fi
\expandafter\ifx\csname citenamefont\endcsname\relax
  \def\citenamefont#1{#1}\fi
\expandafter\ifx\csname url\endcsname\relax
  \def\url#1{\texttt{#1}}\fi
\expandafter\ifx\csname urlprefix\endcsname\relax\def\urlprefix{URL }\fi
\providecommand{\bibinfo}[2]{#2}
\providecommand{\eprint}[2][]{\url{#2}}

\bibitem[{\citenamefont{Crutchfield and McNamara}(1987)}]{Crutchfield1987cs}
\bibinfo{author}{\bibfnamefont{J.}~\bibnamefont{Crutchfield}} \bibnamefont{and}
  \bibinfo{author}{\bibfnamefont{B.}~\bibnamefont{McNamara}},
  \bibinfo{journal}{Comp. sys.} \textbf{\bibinfo{volume}{1}},
  \bibinfo{pages}{417} (\bibinfo{year}{1987}).

\bibitem[{\citenamefont{Kevrekidis et~al.}(2003)\citenamefont{Kevrekidis, Gear,
  Hyman, Kevrekidis, Runborg, and Theodoropoulos}}]{Kevrekidis2003cms}
\bibinfo{author}{\bibfnamefont{I.}~\bibnamefont{Kevrekidis}},
  \bibinfo{author}{\bibfnamefont{C.}~\bibnamefont{Gear}},
  \bibinfo{author}{\bibfnamefont{J.}~\bibnamefont{Hyman}},
  \bibinfo{author}{\bibfnamefont{P.}~\bibnamefont{Kevrekidis}},
  \bibinfo{author}{\bibfnamefont{O.}~\bibnamefont{Runborg}}, \bibnamefont{and}
  \bibinfo{author}{\bibfnamefont{C.}~\bibnamefont{Theodoropoulos}},
  \bibinfo{journal}{Comm. Math. Sci.} \textbf{\bibinfo{volume}{1}},
  \bibinfo{pages}{715} (\bibinfo{year}{2003}).

\bibitem[{\citenamefont{Sugihara et~al.}(2012)\citenamefont{Sugihara, May, Ye,
  Hsieh, Deyle, Fogarty, and Munch}}]{Sugihara2012science}
\bibinfo{author}{\bibfnamefont{G.}~\bibnamefont{Sugihara}},
  \bibinfo{author}{\bibfnamefont{R.}~\bibnamefont{May}},
  \bibinfo{author}{\bibfnamefont{H.}~\bibnamefont{Ye}},
  \bibinfo{author}{\bibfnamefont{C.-h.} \bibnamefont{Hsieh}},
  \bibinfo{author}{\bibfnamefont{E.}~\bibnamefont{Deyle}},
  \bibinfo{author}{\bibfnamefont{M.}~\bibnamefont{Fogarty}}, \bibnamefont{and}
  \bibinfo{author}{\bibfnamefont{S.}~\bibnamefont{Munch}},
  \bibinfo{journal}{\hspace{-.01in}Science\hspace{-.02in}}
  \textbf{\bibinfo{volume}{338}}, \bibinfo{pages}{496} (\bibinfo{year}{2012}).

\bibitem[{\citenamefont{Ye et~al.}(2015)\citenamefont{Ye, Beamish, Glaser,
  Grant, Hsieh, Richards, Schnute, and Sugihara}}]{Ye2015pnas}
\bibinfo{author}{\bibfnamefont{H.}~\bibnamefont{Ye}},
  \bibinfo{author}{\bibfnamefont{R.~J.} \bibnamefont{Beamish}},
  \bibinfo{author}{\bibfnamefont{S.~M.} \bibnamefont{Glaser}},
  \bibinfo{author}{\bibfnamefont{S.~C.} \bibnamefont{Grant}},
  \bibinfo{author}{\bibfnamefont{C.-h.} \bibnamefont{Hsieh}},
  \bibinfo{author}{\bibfnamefont{L.~J.} \bibnamefont{Richards}},
  \bibinfo{author}{\bibfnamefont{J.~T.} \bibnamefont{Schnute}},
  \bibnamefont{and} \bibinfo{author}{\bibfnamefont{G.}~\bibnamefont{Sugihara}},
  \bibinfo{journal}{PNAS} \textbf{\bibinfo{volume}{112}},
  \bibinfo{pages}{E1569} (\bibinfo{year}{2015}).

\bibitem[{\citenamefont{Roberts}(2014)}]{Roberts2014book}
\bibinfo{author}{\bibfnamefont{A.~J.} \bibnamefont{Roberts}},
  \emph{\bibinfo{title}{Model emergent dynamics in complex systems}}
  (\bibinfo{publisher}{SIAM}, \bibinfo{year}{2014}).

\bibitem[{\citenamefont{Schmidt et~al.}(2011)\citenamefont{Schmidt,
  Vallabhajosyula, Jenkins, Hood, Soni, Wikswo, and Lipson}}]{Schmidt2011pb}
\bibinfo{author}{\bibfnamefont{M.}~\bibnamefont{Schmidt}},
  \bibinfo{author}{\bibfnamefont{R.}~\bibnamefont{Vallabhajosyula}},
  \bibinfo{author}{\bibfnamefont{J.}~\bibnamefont{Jenkins}},
  \bibinfo{author}{\bibfnamefont{J.}~\bibnamefont{Hood}},
  \bibinfo{author}{\bibfnamefont{A.}~\bibnamefont{Soni}},
  \bibinfo{author}{\bibfnamefont{J.}~\bibnamefont{Wikswo}}, \bibnamefont{and}
  \bibinfo{author}{\bibfnamefont{H.}~\bibnamefont{Lipson}},
  \bibinfo{journal}{Phys. bio.} \textbf{\bibinfo{volume}{8}},
  \bibinfo{pages}{055011} (\bibinfo{year}{2011}).

\bibitem[{\citenamefont{Daniels and
  Nemenman}(2015{\natexlab{a}})}]{Daniels2015naturecomm}
\bibinfo{author}{\bibfnamefont{B.}~\bibnamefont{Daniels}} \bibnamefont{and}
  \bibinfo{author}{\bibfnamefont{I.}~\bibnamefont{Nemenman}},
  \bibinfo{journal}{Nat. comm.} \textbf{\bibinfo{volume}{6}}
  (\bibinfo{year}{2015}{\natexlab{a}}).

\bibitem[{\citenamefont{Daniels and
  Nemenman}(2015{\natexlab{b}})}]{Daniels2015plosone}
\bibinfo{author}{\bibfnamefont{B.}~\bibnamefont{Daniels}} \bibnamefont{and}
  \bibinfo{author}{\bibfnamefont{I.}~\bibnamefont{Nemenman}},
  \bibinfo{journal}{PloS one} \textbf{\bibinfo{volume}{10}},
  \bibinfo{pages}{e0119821} (\bibinfo{year}{2015}{\natexlab{b}}).

\bibitem[{\citenamefont{Bongard and Lipson}(2007)}]{Bongard2007pnas}
\bibinfo{author}{\bibfnamefont{J.}~\bibnamefont{Bongard}} \bibnamefont{and}
  \bibinfo{author}{\bibfnamefont{H.}~\bibnamefont{Lipson}},
  \bibinfo{journal}{PNAS} \textbf{\bibinfo{volume}{104}}, \bibinfo{pages}{9943}
  (\bibinfo{year}{2007}).

\bibitem[{\citenamefont{Schmidt and Lipson}(2009)}]{Schmidt2009science}
\bibinfo{author}{\bibfnamefont{M.}~\bibnamefont{Schmidt}} \bibnamefont{and}
  \bibinfo{author}{\bibfnamefont{H.}~\bibnamefont{Lipson}},
  \bibinfo{journal}{Science} \textbf{\bibinfo{volume}{324}},
  \bibinfo{pages}{81} (\bibinfo{year}{2009}).

\bibitem[{\citenamefont{Tibshirani}(1996)}]{Tibshirani1996lasso}
\bibinfo{author}{\bibfnamefont{R.}~\bibnamefont{Tibshirani}},
  \bibinfo{journal}{J. Roy. Stat. Soc. B} p. \bibinfo{pages}{267}
  (\bibinfo{year}{1996}).

\bibitem[{\citenamefont{Brunton et~al.}(2016)\citenamefont{Brunton, Proctor,
  and Kutz}}]{Brunton2016}
\bibinfo{author}{\bibfnamefont{S.~L.} \bibnamefont{Brunton}},
  \bibinfo{author}{\bibfnamefont{J.~L.} \bibnamefont{Proctor}},
  \bibnamefont{and} \bibinfo{author}{\bibfnamefont{J.~N.} \bibnamefont{Kutz}},
  \bibinfo{journal}{PNAS} \textbf{\bibinfo{volume}{113}}, \bibinfo{pages}{3932}
  (\bibinfo{year}{2016}).

\bibitem[{\citenamefont{Mangan et~al.}(2016)\citenamefont{Mangan, Brunton,
  Proctor, and Kutz}}]{Mangan2016}
\bibinfo{author}{\bibfnamefont{N.~M.} \bibnamefont{Mangan}},
  \bibinfo{author}{\bibfnamefont{S.~L.} \bibnamefont{Brunton}},
  \bibinfo{author}{\bibfnamefont{J.~L.} \bibnamefont{Proctor}},
  \bibnamefont{and} \bibinfo{author}{\bibfnamefont{J.~N.} \bibnamefont{Kutz}},
  \bibinfo{journal}{ArXiv e-prints arXiv:1605.08368}  (\bibinfo{year}{2016}).

\bibitem[{\citenamefont{Holmes et~al.}(2012)\citenamefont{Holmes, Lumley,
  Berkooz, and Rowley}}]{HLBR_turb}
\bibinfo{author}{\bibfnamefont{P.}~\bibnamefont{Holmes}},
  \bibinfo{author}{\bibfnamefont{J.}~\bibnamefont{Lumley}},
  \bibinfo{author}{\bibfnamefont{G.}~\bibnamefont{Berkooz}}, \bibnamefont{and}
  \bibinfo{author}{\bibfnamefont{C.}~\bibnamefont{Rowley}},
  \emph{\bibinfo{title}{Turbulence, coherent structures, dynamical systems and
  symmetry}} (\bibinfo{publisher}{Cambridge}, \bibinfo{address}{Cambridge,
  England}, \bibinfo{year}{2012}), \bibinfo{edition}{2nd} ed.

\bibitem[{\citenamefont{Einstein}(1905)}]{Einstein1905brownian}
\bibinfo{author}{\bibfnamefont{A.}~\bibnamefont{Einstein}},
  \bibinfo{journal}{Ann. der Physik} \textbf{\bibinfo{volume}{17}},
  \bibinfo{pages}{549} (\bibinfo{year}{1905}).

\end{thebibliography}

\end{document}